\documentclass[a4paper,usenatbib]{mnras}

\usepackage{amsmath}
\usepackage{amssymb}
\usepackage{newtxtext,newtxmath}
\usepackage[T1]{fontenc}
\usepackage{ae,aecompl}
\usepackage{graphicx}
\usepackage{times}
\usepackage{xcolor}

\title[EAGLE infrared luminosity and dust mass functions]{Infrared luminosity functions and dust mass functions in the EAGLE simulation}

\author[M.~Baes et al.]{%
Maarten Baes,$^1$
Ana Tr\v{c}ka,$^1$
Peter Camps,$^1$
James Trayford,$^2$
Antonios Katsianis,$^{3,4}$\newauthor
Lucia Marchetti,$^{5,6,7}$
Tom Theuns,$^8$
Mattia Vaccari,$^{6,7}$ and
Bert Vandenbroucke$^1$
\vspace*{1ex} \\
$^1$Sterrenkundig Observatorium, Universiteit Gent, Krijgslaan 281 S9, B-9000 Gent, Belgium \\
$^2$Leiden Observatory, Leiden University, P.O. Box 9513, NL-2300 RA Leiden, The Netherlands \\
$^3$Tsung-Dao Lee Institute, Shanghai Jiao Tong University, Shanghai 200240, China \\
$^4$Department of Astronomy, Shanghai Key Laboratory for Particle Physics and Cosmology, Shanghai Jiao Tong University, Shanghai 200240, China \\
$^5$Department of Astronomy, University of Cape Town, Private Bag X3, Rondebosch 7701, Cape Town, South Africa \\
$^6$Department of Physics and Astronomy, University of the Western Cape, Private Bag X17, 7535 Bellville, Cape Town, South Africa \\
$^7$INAF - Istituto di Radioastronomia, via Gobetti 101, 40129 Bologna, Italy \\
$^8$Institute for Computational Cosmology, Department of Physics, University of Durham, South Road, Durham DH1 3LE, UK
}

\date{Accepted 2020 April 5. Received 2020 March 16; in original form 2019 December 20}

\pubyear{2019}

\begin{document}
\label{firstpage}
\pagerange{\pageref{firstpage}--\pageref{lastpage}}

\maketitle

\begin{abstract}
We present infrared luminosity functions and dust mass functions for the EAGLE cosmological simulation, based on synthetic multi-wavelength observations generated with the SKIRT radiative transfer code. In the local Universe, we reproduce the observed infrared luminosity and dust mass functions very well. Some minor discrepancies are encountered, mainly in the high luminosity regime, where the EAGLE-SKIRT luminosity functions mildly but systematically underestimate the observed ones. The agreement between the EAGLE-SKIRT infrared luminosity functions and the observed ones gradually worsens with increasing lookback time. Fitting modified Schechter functions to the EAGLE-SKIRT luminosity and dust mass functions at different redshifts up to $z=1$, we find that the evolution is compatible with pure luminosity/mass evolution. The evolution is relatively mild: within this redshift range, we find an evolution of $L_{\star,250}\propto(1+z)^{1.68}$, $L_{\star,\text{TIR}}\propto(1+z)^{2.51}$ and $M_{\star,\text{dust}}\propto(1+z)^{0.83}$ for the characteristic luminosity/mass. For the luminosity/mass density we find $\varepsilon_{250}\propto(1+z)^{1.62}$, $\varepsilon_{\text{TIR}}\propto(1+z)^{2.35}$ and $\rho_{\text{dust}}\propto(1+z)^{0.80}$, respectively. The mild evolution of the dust mass density is in relatively good agreement with observations, but the slow evolution of the infrared luminosity underestimates the observed luminosity evolution significantly. We argue that these differences can be attributed to increasing limitations in the radiative transfer treatment due to increasingly poorer resolution, combined with a slower than observed evolution of the SFR density in the EAGLE simulation and the lack of AGN emission in our EAGLE-SKIRT post-processing recipe.
\end{abstract}

\begin{keywords}
galaxies: evolution -- cosmology: observations -- radiative transfer -- hydrodynamics
\end{keywords}

\newpage
\newpage
\section{Introduction}

Because of the enormous astrophysical complexity and the vast range of scales at play, galaxy formation and evolution studies rely more and more on complex numerical models. Cosmological hydrodynamical simulations form one of the leading techniques in this field \citep{2015ARA&A..53...51S}. Modern hydrodynamical simulation suites such as Illustris \citep{2014MNRAS.444.1518V}, Horizon-AGN \citep{2016MNRAS.463.3948D, 2017MNRAS.467.4739K}, EAGLE \citep{2015MNRAS.450.1937C, 2015MNRAS.446..521S}, MassiveBlack \citep{2015MNRAS.450.1349K}, MUFASA \citep{2016MNRAS.462.3265D, 2019MNRAS.486.2827D} and IllustrisTNG \citep{2018MNRAS.473.4077P, 2019MNRAS.490.3196P} have become fundamental tools in our endeavour to understand galaxy formation and evolution. For a general overview, see \citet{2015ARA&A..53...51S} or \citet{2020NatRP...2...42V}.

In order to test the validity and predictive power of cosmological hydrodynamical simulations, they need to be confronted with observational data. While this comparison can be done in physical space, this brings along a number of disadvantages. Indeed, while the intrinsic properties, such as stellar masses, gas metallicities or star formation rates, of simulated galaxies can directly be extracted from the simulation data, they are not directly measurable for observed galaxies. Instead, they have to be inferred based on models, which are characterised by explicit or implicit assumptions, biases and simplifications, for example on the star formation history or the effect of dust attenuation. Even for `simple' intrinsic properties such as stellar masses or star formation rates, different assumptions can lead to significantly varying results \citep[e.g.,][]{2013ARA&A..51..393C, 2013MNRAS.435...87M, 2016PASA...33...29K, 2019ApJ...879...11K, 2020MNRAS.492.5592K, 2019A&A...621A..51H}. More complex diagnostics such as the star formation history are evidently even harder to infer \citep{2015MNRAS.453.1597S, 2019ApJ...876....3L}. 

An alternative approach, which is gaining more popularity, is to compare simulations and observations directly in the observer's frame \citep[e.g.,][]{2010MNRAS.403...17J, 2010MNRAS.407L..41S, 2012MNRAS.424..951H, 2015MNRAS.447.2753T, 2016MNRAS.462.2046G, 2018MNRAS.479..917G, 2017MNRAS.470..771T, 2017MNRAS.469.3775G, 2019MNRAS.483.4140R, 2020MNRAS.492.5167V}. This direct modelling approach involves the creation of synthetic observables, in which, ideally, all the necessary physical recipes and instrumental characteristics are included. One crucial aspect when creating realistic synthetic observables is the presence of interstellar dust. Cosmic dust affects the observations of galaxies over the entire UV--submm spectrum: it is very efficient at absorbing and scattering UV and optical radiation, and dominates the entire mid-infrared to submm wavelength range through direct thermal emission. Properly taking into account the effects of interstellar dust on the observed properties of galaxies requires dust radiative transfer calculations. Such calculations are nowadays perfectly doable, even for complex 3D geometries \citep[for a review, see][]{2013ARA&A..51...63S}. 

In \citet{2016MNRAS.462.1057C, 2018ApJS..234...20C} we used a physically motivated recipe to include interstellar dust in the EAGLE simulations. In general, this so-called EAGLE-SKIRT post-processing recipe yields results that agree very well with observations, including optical colours and the stellar-mass versus colour diagram \citep{2017MNRAS.470..771T}, the cosmic spectral energy distribution (CSED) in the Local Universe \citep{2019MNRAS.484.4069B}, dust scaling relations for local galaxies \citep{Trcka2019}, and non-parametric morphology statistics \citep{2019arXiv190810936B}. However, some tensions between the EAGLE-SKIRT synthetic data and observations were found as well. The simulated galaxies do not show the same dependence of attenuation on inclination as found in observations \citep{2017MNRAS.470..771T}, the average UV attenuation in local galaxies tends to be underestimated \citep{Trcka2019}, the number of submm galaxies with high star formation rates at high redshifts underestimates the observed number \citep{2019MNRAS.488.2440M}, and we fail to reproduce the strong evolution of the CSED with increasing redshift, particular at far-infrared wavelengths \citep{2019MNRAS.484.4069B}.

In this paper, we focus on the comparison of EAGLE-SKIRT luminosity functions and observations at infrared wavelengths. Since the 1980s, luminosity functions have been measured in different infrared and submm bands, both in the local Universe and out to intermediate redshifts \citep[e.g.,][]{1990MNRAS.242..318S, 2005ApJ...632..169L, 2006MNRAS.370.1159B, 2007ApJ...663..218M, 2010A&A...518L..10D, 2010A&A...515A...8R, 2010A&A...518L..20V, 2010A&A...518L..23E, 2013MNRAS.432...23G, 2018MNRAS.473.3507E, 2013MNRAS.429.1309N, 2016MNRAS.456.1999M}. As the thermal emission by interstellar dust completely dominates the emission of galaxies at infrared and submm wavelengths, a comparison of EAGLE-SKIRT infrared luminosity functions to observed ones provides a strong test for the validity of our dust post-processing recipes. In particular, as the luminosity functions contain more fine-grained information than the global CSED, they can provide useful information to identify the shortcomings in our recipes, and guide the way to more advanced algorithms.

This paper is built up as follows. In Sect.~{\ref{EAGLE.sec}} we briefly describe the EAGLE simulations and the EAGLE-SKIRT database we use for our study. In Sect.~{\ref{LF.sec}} we present luminosity functions for the EAGLE simulation in the local Universe for different infrared/submm bands, and compare them to observations. In Sect.~{\ref{Derived.sec}} we consider the total infrared luminosity function, the dust mass function, and the infrared cosmic spectral energy distribution. In Sect.~{\ref{LFevo.sec}} we investigate the evolution of the EAGLE-SKIRT luminosity functions and dust mass functions, and compare them to observations. In Sect.~{\ref{Discussion.sec}} we discuss the implications of our results, and in Sect.~{\ref{Summary.sec}} we summarise.

\section{The data}
\label{EAGLE.sec}

EAGLE \citep[Evolution and Assembly of GaLaxies and their Environments;][]{2015MNRAS.450.1937C, 2015MNRAS.446..521S} is a suite of cosmological hydrodynamical simulations, performed in cubic boxes with a range of sizes. The simulations have been calibrated to reproduce the local stellar mass function, the galaxy-central black hole mass relation, and the galaxy mass-size relation, and have been compared to many other diagnostics \citep[e.g.,][]{2015MNRAS.446..521S, 2015MNRAS.452.3815L, 2017MNRAS.464.3850L, 2016MNRAS.462..190R, 2017MNRAS.464.4204C, 2017MNRAS.465..722F, 2017MNRAS.472..919K}. In this paper, we will focus on the RefL0100N1504 and RecalL0025N0752 simulations, hereafter referred to as Ref-100 and Recal-25 respectively. The former run is the reference EAGLE simulation, covering the largest volume of all EAGLE runs (100 cMpc on the side). The latter covers a smaller volume (25 cMpc on the side), but has an eight times better mass resolution. The subgrid parameters of this latter run have been recalibrated to ensure weak convergence \citep[see][for details]{2015MNRAS.450.1937C, 2015MNRAS.446..521S}.

\citet{2016MNRAS.462.1057C} introduced a framework to incorporate interstellar dust in the EAGLE galaxies. This framework consists of a resampling procedure for star forming particles, the use of subgrid templates for dusty star forming regions, the inclusion of diffuse dust based on the distribution of metals in the gas phase, and full 3D dust radiative transfer modelling using SKIRT \citep{2003MNRAS.343.1081B, 2011ApJS..196...22B, 2015A&C.....9...20C}. The parameters in the post-processing scheme were calibrated to reproduce the observed submm colours and dust scaling relations of nearby Herschel Reference Survey (HRS) galaxies \citep{2012A&A...540A..54B, 2012A&A...540A..52C, 2014MNRAS.440..942C}. \citet{2018ApJS..234...20C} used this framework to generate synthetic observations for six different EAGLE simulations, including the Ref-100 and Recal-25 simulations. The resulting EAGLE-SKIRT database contains synthetic UV to submm flux densities and intrinsic luminosities for all galaxies in the simulations with at least 250 dust containing particles, and with stellar masses above $10^{8.5}~M_\odot$. Synthetic observables are available for 23 redshift slices, ranging from $z=0$ to $z=6$. These synthetic data, which we refer to as the EAGLE-SKIRT data, are available from the public EAGLE galaxy database\footnote{\href{http://www.eaglesim.org/database.php}{http://www.eaglesim.org/database.php}} \citep{2016A&C....15...72M}. 

\section{Monochromatic luminosity functions}
\label{LF.sec}

In this Section we show monochromatic EAGLE-SKIRT luminosity functions in various infrared broadband filters, and compare them to observational data. We mainly focus on the redshift range $z\leqslant0.2$, for which observational data is available in several bands.

\subsection{Calculation of the luminosity functions}

To calculate the EAGLE-SKIRT luminosity functions, we extract the infrared luminosities for all galaxies from the EAGLE-SKIRT database. For each of the two EAGLE volumes considered, and for each infrared broadband and redshift, the luminosity function is calculated by simply binning the number of sources per logarithmic bin in luminosity $L=\nu L_\nu$, and dividing by the co-moving volume of the simulation. We subsequently average the luminosity functions at the lowest three redshifts bins ($z=0$, $0.1$ and $0.18$). Finally, we combine the luminosity functions of the Ref-100 and Recal-25 simulations, based on the number of sources in each bin. When the number of Recal-25 sources in the bin is larger than 20, we use that estimate for the luminosity function. When it is lower than 10, we use the Ref-100 estimate of the luminosity function. When the number of Recal-25 sources in a bin is between 10 and 20, we take the logarithmic mean of the Recal-25 and Ref-100 estimates of the luminosity function as our final estimate. The rationale for this procedure is that the Recal-25 simulation, with its better mass resolution, provides the best constraints at low and intermediate luminosities. At the high-luminosity end, the Ref-100 simulation, with its larger volume, provides better statistics and hence better constraints on the luminosity function. In the overlap region, the estimates of both simulations agree very well. The error bars on the EAGLE-SKIRT luminosity functions reflect the 1$\sigma$ Poisson uncertainties.

We also fit parametric functions to the discrete luminosity functions as calculated above. While various alternative options could be used \citep[e.g.,][]{1986MNRAS.219..687L, 1993ApJS...89....1R, 2013MNRAS.428..291P}, we follow the standard practice in the infrared astronomy community \citep{1990MNRAS.242..318S, 2005ApJ...632..169L, 2010A&A...515A...8R, 2013MNRAS.428..291P, 2016MNRAS.456.1999M}, and use the modified Schechter function,
\begin{multline}
\Phi(L)\,{\text{d}}\log L \\
= \Phi^\star \left(\frac{L}{L^\star}\right)^{1-\alpha}
\exp\left[-\frac{1}{2\sigma^2}\log^2\left(1+\frac{L}{L^\star}\right)\right] {\text{d}}\log L.
\end{multline}
This function behaves as a power law in the low-luminosity regime ($L\ll L^\star$) and as a log-normal function for $L\gg L^\star$. It is characterised by four parameters: the characteristic density $\Phi^\star$ is a normalisation factor defining the overall galaxy density, $L^\star$ is the characteristic luminosity, $\alpha$ represents the faint-end slope of the luminosity function, and $\sigma$ characterises the width of the lognormal distribution. 

\subsection{Submm luminosity functions}
\label{LF-SPIRE.sec}

\begin{figure*}
\centering
\includegraphics[width=\textwidth]{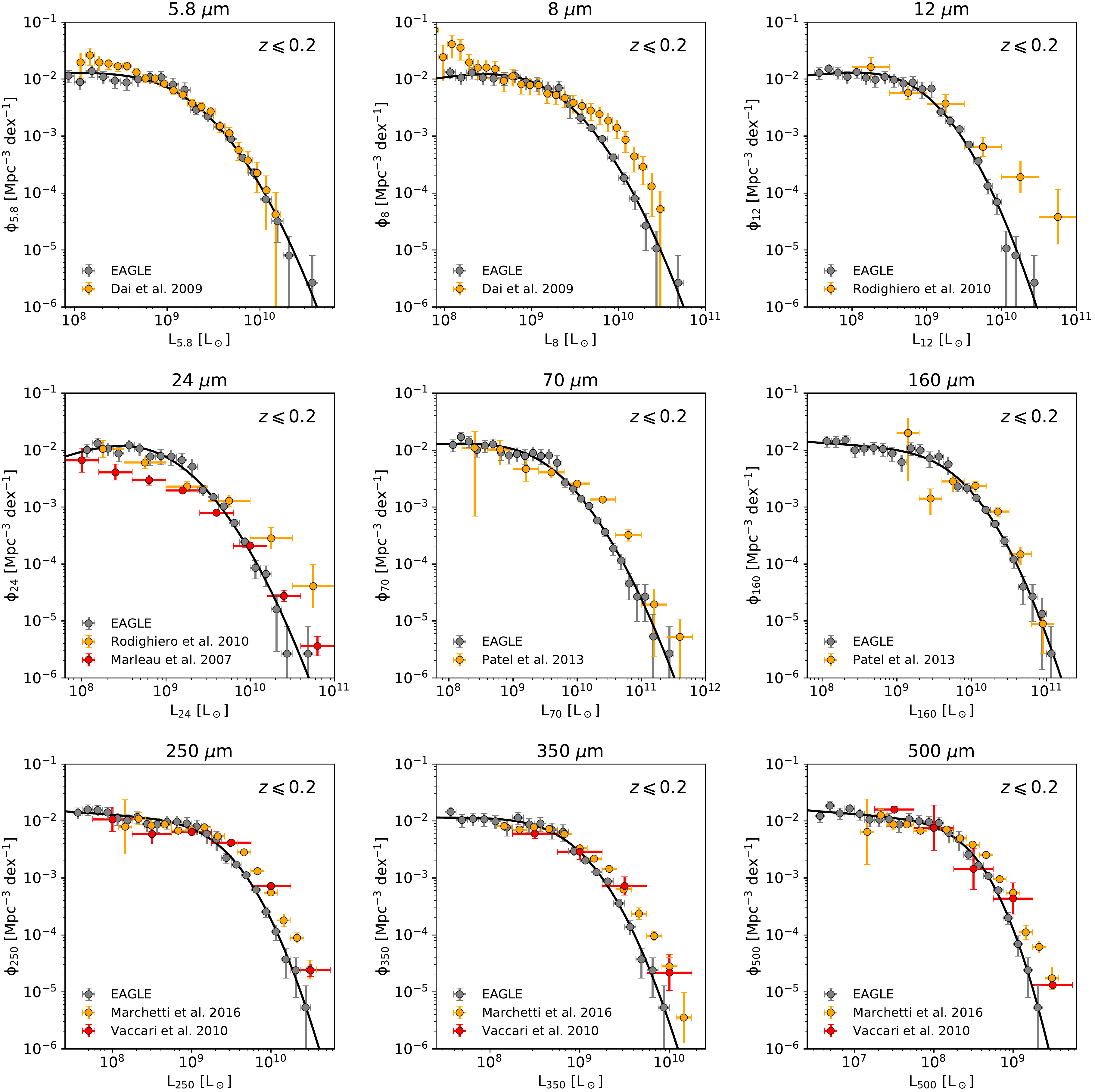}
\caption{Luminosity functions for the EAGLE simulations for $z\leqslant0.2$ in different infrared and submm broadband filters. In each panel, grey dots represent the EAGLE-SKIRT luminosity function as obtained from the EAGLE-SKIRT database, and the solid line is the best modified Schechter fit to these data points. The data points with error bars represent observed luminosity functions from different sources, as indicated in the bottom left corner of each panel.} 
\label{EAGLE-LF-z00-02.fig}
\end{figure*}

\begin{figure*}
\centering
\includegraphics[width=\textwidth]{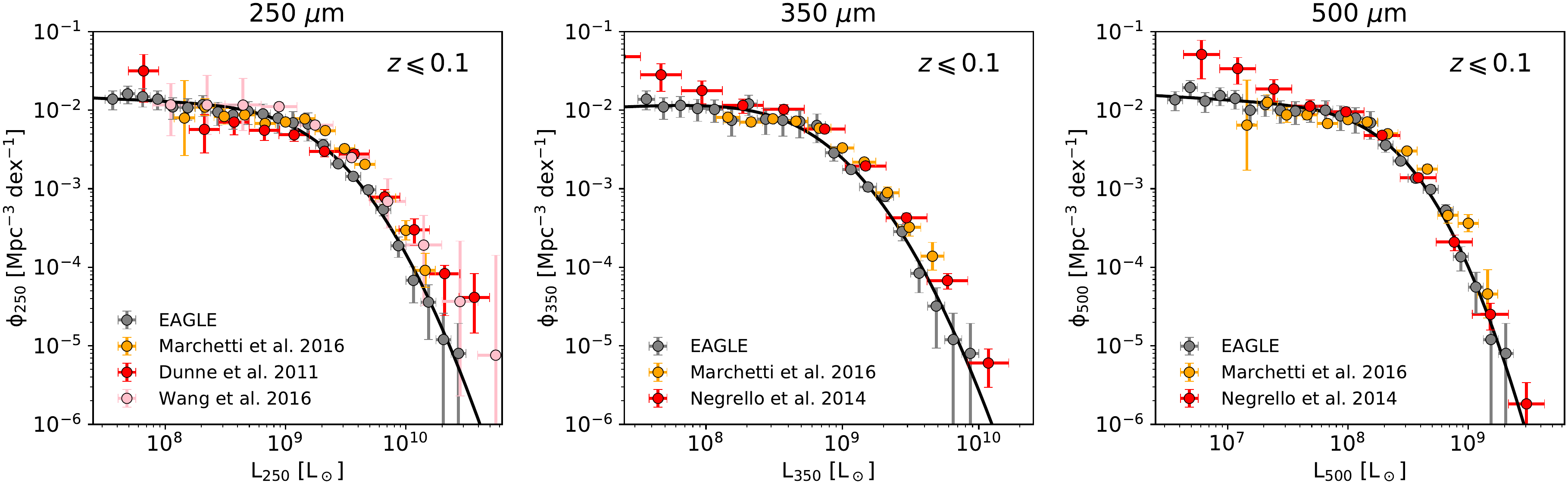}
\includegraphics[width=0.7\textwidth]{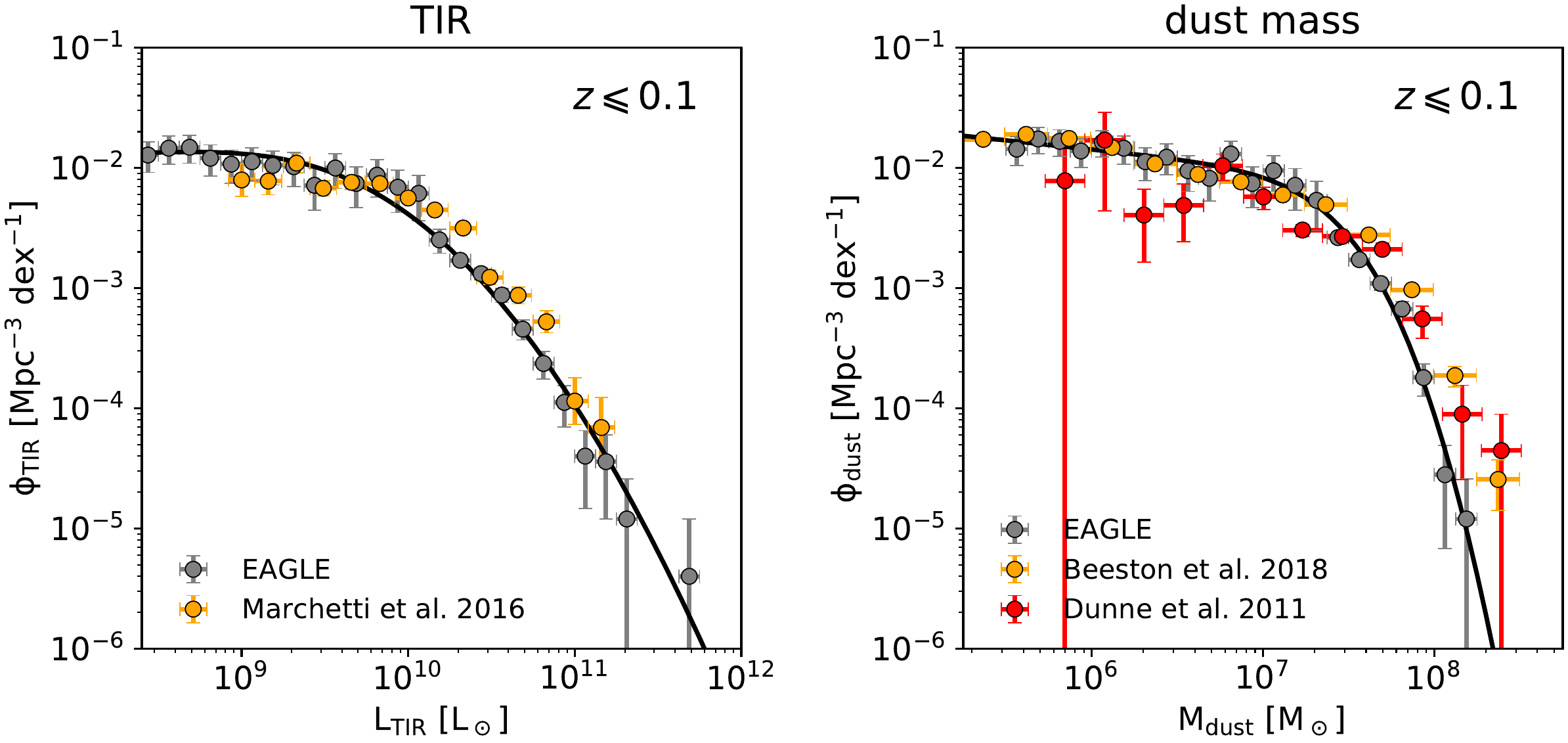}
\caption{Same as Fig.~{\ref{EAGLE-LF-z00-02.fig}}, but now restricted to the redshift range $z\leqslant0.1$. Shown are luminosity functions in the three Herschel SPIRE bands, the total infrared luminosity function and the dust mass function.} 
\label{EAGLE-LF-z00-01.fig}
\end{figure*}

The three panels on the bottom row in Fig.~{\ref{EAGLE-LF-z00-02.fig}} show the EAGLE-SKIRT luminosity function in the three Herschel SPIRE bands (250, 350 and 500 $\mu$m). We compare the EAGLE-SKIRT results to luminosity functions obtained by \citet{2010A&A...518L..20V} and \citet{2016MNRAS.456.1999M}, both based on data obtained in the frame of the HerMES survey \citep{2012MNRAS.424.1614O}. Both data sets are in clear agreement with each other. At low luminosities, the agreement between the EAGLE-SKIRT and HerMES luminosity functions in the SPIRE bands is excellent. At high luminosities, however, we note a systematic difference, in the sense that the EAGLE-SKIRT luminosity functions gradually start to underestimate the HerMES luminosity functions for $L\gtrsim L^\star$. The same systematic behaviour is seen for the three SPIRE bands. 

At first sight, this systematic difference is somewhat unexpected, as our EAGLE-SKIRT post-processing recipe is primarily calibrated based on SPIRE data \citep{2016MNRAS.462.1057C}. However, the HRS galaxies used in the calibration process are all normal late-type star-forming galaxies, which populate the low-luminosity side of the luminosity function \citep{2014A&A...566A..70A}. This explains the excellent agreement between EAGLE-SKIRT and HerMES in his regime, but also leaves the high-luminosity tail unconstrained. 

We believe there are three explanations that can jointly explain this systematic difference at the high-luminosity tail. Firstly, the high-luminosity tail of the SPIRE luminosity functions can be affected by submm sources that are not present in our EAGLE simulation. For example, one of the most luminous SPIRE sources in the local Universe is M87, whose submm emission is not due to thermal dust emission but to synchrotron emission \citep{2010A&A...518L..53B, 2010A&A...518L..61B}. Gravitational lensing can also boost the luminosity function in the high-luminosity tail \citep{2007MNRAS.377.1557N, 2010Sci...330..800N, 2012ApJ...749...65G}. 

A second factor could be due to AGN emission. The AGN emission peaks at mid-infrared rather than submm wavelengths \citep{2010A&A...518L..33H, 2015A&A...576A..10C, 2017MNRAS.466.3161S}, but AGNs, and luminous QSOs in particular, can also contribute substantially to the emission in the Herschel SPIRE bands \citep[e.g.,][]{2016MNRAS.459..257S, 2019arXiv190804795K}. Supermassive black hole growth and feedback are crucial ingredients of the EAGLE simulations \citep{2016MNRAS.462..190R, 2017MNRAS.468.3395M}, but our post-processing does not (yet) take into account AGN emission. A coupling to the SKIRTOR library of clumpy AGN torus models \citep{2012MNRAS.420.2756S, 2016MNRAS.458.2288S}, calculated in a self-consistent way with the same radiative transfer code, is foreseen for the near future.

A third factor that could potentially contribute to this systematic difference is source confusion. Due to the large beam of SPIRE, source confusion and blending issues can lead to an overestimation of source counts and luminosity functions, as convincingly demonstrated by \citet{2019A&A...624A..98W}. We note, however, that the HerMES luminosity functions of \citet{2016MNRAS.456.1999M} were based on XID catalogues \citep{2010MNRAS.409...48R}, which use deeper 24 $\mu$m detections as prior for the SPIRE source extraction. This should in principle limit the problems due to blending, although some minor effect might still be at play, particularly at the longer SPIRE wavelengths.

Finally, evolution effects are probably already at play, even though we only consider the relatively narrow redshift range $z\leqslant0.2$. Several studies have suggested that the SPIRE luminosity function of galaxies evolves very rapidly at low redshifts \citep{2010A&A...518L..10D, 2011MNRAS.417.1510D, 2016MNRAS.456.1999M, 2016A&A...592L...5W}. The top row of Fig.~{\ref{EAGLE-LF-z00-01.fig}} shows similar SPIRE luminosity functions as the bottom row of Fig.~{\ref{EAGLE-LF-z00-02.fig}}, but now restricted to the redshift range $z\leqslant0.1$.\footnote{The difference between the SPIRE 250~$\mu$m EAGLE-SKIRT luminosity functions corresponding to $z\leqslant0.1$ and $z\leqslant0.2$ is about 0.07 dex at $L_{250} = 10^9~L_\odot$, and 0.23 dex at $L_{250} = 10^{10}~L_\odot$. We discuss the evolution of the infrared luminosity functions in more detail in Sect.~{\ref{LFevo.sec}}.}
Again we show HerMES data from \citet{2016MNRAS.456.1999M}, but also the 250~$\mu$m luminosity function by \citet{2011MNRAS.417.1510D} obtained for the H-ATLAS survey \citep{2010PASP..122..499E}, the 250~$\mu$m luminosity function derived by \citet{2016A&A...592L...5W} for the Herschel Stripe~82 Survey \citep{2014ApJS..210...22V}, and the 350 and 500 $\mu$m luminosity functions as derived by \citet{2013MNRAS.429.1309N}, based on data from the Planck Early Release Compact Source Catalogue \citep{2011A&A...536A...7P}.\footnote{The Planck luminosities have been converted from 550 $\mu$m to 500 $\mu$m assuming a modified blackbody spectrum with an effective emissivity index $\beta=1.8$, appropriate in the local Universe \citep{2011A&A...536A..21P, 2013MNRAS.436.2435S, 2014MNRAS.440..942C}.} There is some tension between the Herschel- and Planck-based luminosity functions at both 350 and 500 $\mu$m: \citet{2013MNRAS.429.1309N} find a steep increase of the luminosity function in the lowest luminosity bins, whereas the HerMES luminosity functions remain roughly flat. As argued by \citet{2016MNRAS.456.1999M}, this conspicuous increase is likely due to the contamination by the Local Supercluster or the Virgo Cluster. Compared to the $z\leqslant0.2$ luminosity functions shown in Fig.~{\ref{EAGLE-LF-z00-02.fig}}, the systematic difference between EAGLE-SKIRT and observations is reduced. At 500 $\mu$m, it disappears almost completely. This hints that evolution might indeed play a prominent role in this discrepancy. 

\subsection{Mid- and far-infrared luminosity functions}
\label{LF-infrared.sec}

On the top two rows of Fig.~{\ref{EAGLE-LF-z00-02.fig}}, we present the EAGLE-SKIRT $z\leqslant0.2$ luminosity functions in a number of mid- and far-infrared bands, and compare them to observational results corresponding to the same redshift range. Note that data in this wavelength range were not involved in the calibration of the EAGLE-SKIRT post-processing recipe of \citet{2016MNRAS.462.1057C}, so in theory these luminosity functions are more stringent tests for the EAGLE-SKIRT results.

For the Spitzer IRAC bands at 5.8 and 8~$\mu$m, we compare the EAGLE-SKIRT results to the luminosity functions obtained by \citet{2009ApJ...697..506D} in the frame of the AGN and Galaxy Evolution Survey \citep[AGES:][]{2012ApJS..200....8K}. At 5.8~$\mu$m, the agreement between EAGLE-SKIRT and AGES is excellent, except in the low-luminosity regime where the EAGLE-SKIRT results slightly underestimate the observations. At 8~$\mu$m, the agreement is less convincing: simulation and observations agree around $L_8\sim10^9~L_\odot$, but the EAGLE-SKIRT luminosity function underestimates the observed luminosity function at higher luminosities. It must be mentioned that the AGES 8~$\mu$m luminosity function has a particular shape, with a conspicuous excess for $L_8\gtrsim10^{10}~L_\odot$. \citet{2009ApJ...697..506D} explain this particular shape as the result of very different luminosity functions for early-type galaxies and late-type galaxies, with the strong PAH emission from the latter being absent in the former. In our SKIRT post-processing, we have assumed a single uniform dust mixture \citep{2004ApJS..152..211Z} within and among all galaxies, and it is therefore not surprising that we do not reproduce the detailed PAH emission of observed galaxies. In our study of the EAGLE-SKIRT CSED \citep{2019MNRAS.484.4069B}, we also see that the global emission in the IRAC 8 $\mu$m band underestimates the observed emission by about 0.2 dex for $z\sim0.05$.

At 12 and 24~$\mu$m, we compare our EAGLE-SKIRT luminosity functions to observational data obtained by \citet{2010A&A...515A...8R}, based on a combination of data from deep Spitzer surveys of the VIMOS VLT Deep Survey (VVDS-SWIRE) and GOODS fields. At low luminosities, the agreement between EAGLE-SKIRT and the observations is satisfactory, but a strong difference is found at high luminosities. In particular, the EAGLE-SKIRT luminosity function strongly underestimates the high-luminosity tail at both wavelengths, with difference exceeding an order of magnitude in the highest luminosity bins. This is essentially the same problem as found for the submm luminosity functions.

We identify four different reasons that can contribute to explain this discrepancy. Firstly, the luminosity at mid-infrared wavelengths mainly originates from star forming regions, which are below the resolution limit for the EAGLE simulations, and hence also for the SKIRT radiative transfer post-processing. This resolution issue is handled using a subgrid approach, first employed by \citet{2010MNRAS.403...17J}: we represent the star-forming regions using a template SEDs from the MAPPINGS library \citep{2008ApJS..176..438G}. Because the MIR range was not directly involved in the original calibration of the subgrid parameters \citep{2016MNRAS.462.1057C, 2018ApJS..234...20C}, some MAPPINGS parameters are relatively unconstrained, resulting in relatively poor agreement with observational data \citep{2019MNRAS.484.4069B, Trcka2019}.

A second factor is the lack of AGN emission in our post-processing routine, already mentioned in previous subsection. AGN emission typically peaks at mid-infrared wavelengths, so the missing contribution by AGNs could be a significant factor in the discrepancy between the EAGLE-SKIRT and observed luminosity functions at 12 and 24~$\mu$m.

Thirdly, the same cosmic evolution effects that we believe were partly responsible for the discrepancies in the SPIRE bands are probably also at play here. A strong evolution of the mid- and far-infrared luminosity function has become evident from several observational studies \citep[e.g.,][]{2005ApJ...632..169L, 2006MNRAS.370.1159B, 2007ApJ...660...97C, 2010A&A...515A...8R}, and we will come back to evolutionary effects in Sect.~{\ref{LFevo.sec}}.

A final possibility is that the luminosity functions as measured by \citet{2010A&A...515A...8R} are overestimated, particularly at the high luminosity end. In the central left panel of Fig.~{\ref{EAGLE-LF-z00-02.fig}} we also show the 24~$\mu$m luminosity function for $z<0.25$ as obtained by \citet{2007ApJ...663..218M} in the frame of the Spitzer Extragalactic First Look Survey \citep[FLS:][]{2006AJ....131.2859F}. While the redshift range is not exactly equal, the two observed luminosity functions differ considerably, particularly in the high-luminosity regime. In fact, the EAGLE-SKIRT 24~$\mu$m luminosity function agrees rather well with the FLS luminosity function for $L_{24}\gtrsim10^{10}~L_\odot$.

Finally, the central middle and right panels of Fig.~{\ref{EAGLE-LF-z00-02.fig}} show the EAGLE-SKIRT luminosity functions in the MIPS 70 and 160 $\mu$m bands, as well as the observed SWIRE luminosity functions taken from \citet{2013MNRAS.428..291P}. The agreement with the EAGLE-SKIRT data is, overall, very satisfactory. At 70~$\mu$m, the EAGLE-SKIRT luminosity function only slightly underestimates the observed luminosity function in the high luminosity regime ($L_{70}\gtrsim10^{10}~L_\odot$).

\section{Derived properties}
\label{Derived.sec}

\begin{figure*}
\centering
\includegraphics[width=0.7\textwidth]{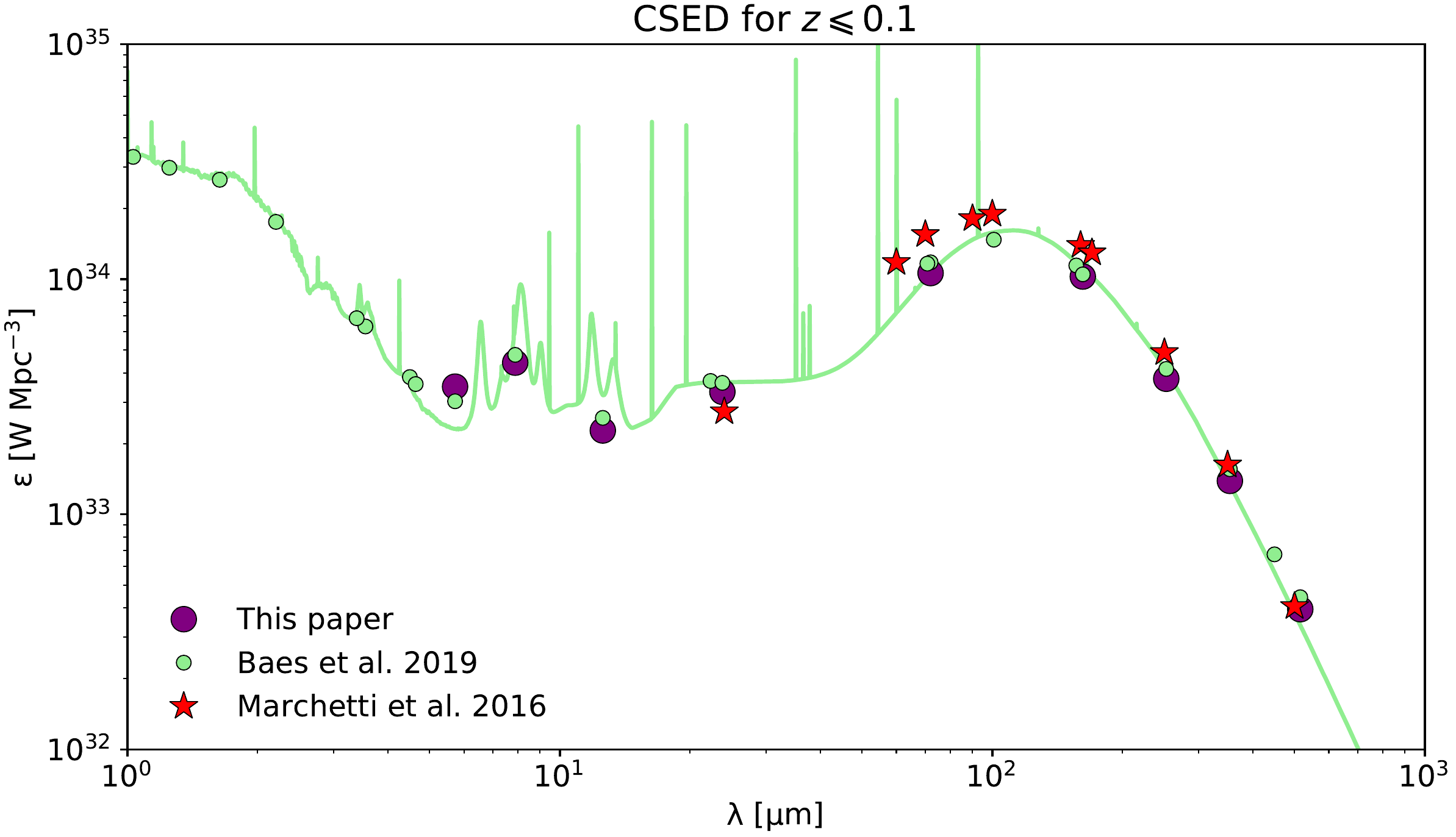}
\caption{The infrared cosmic spectral energy distribution in the Local Universe. Purple dots are calculated by integrating the EAGLE-SKIRT luminosity functions as calculated in this paper. The green dots  correspond to the EAGLE-SKIRT CSED data calculated by \citet{2019MNRAS.484.4069B}, and the green line is the best-fitting CIGALE \citep{2019A&A...622A.103B} SED model through these data points. The red stars are the HerMES CSED data points from \citet{2016MNRAS.456.1999M}.} 
\label{CSED-z00-01.fig}
\end{figure*}

\subsection{The total infrared luminosity function}
\label{totLF.sec}

Apart from monochromatic luminosity functions corresponding to individual infrared bands, it is interesting to look at the total or bolometric infrared luminosity function. Compared to monochromatic luminosities, the total infrared luminosity is a physical quantity that has a more direct meaning: it corresponds to the total amount of stellar radiation that has been absorbed and re-emitted by the dusty interstellar medium (plus the possible contribution from an AGN). The total infrared emission is one of the most popular tracers for the star formation rate, either by itself or in combination with UV or H$\alpha$ luminosities \citep{2009ApJ...703.1672K, 2011ApJ...741..124H, 2011ApJ...737...67M}. When combined with other observables such as the UV luminosity or the total stellar luminosity, it is an important diagnostic for the dust content or attenuation in galaxies \citep{1996A&A...306...61B, 2012A&A...539A.145B, 2016A&A...586A..13V}.

For the galaxies in the EAGLE simulation, the total infrared luminosity can in principle directly be measured from the fully sampled SKIRT spectral energy distribution. However, in order to mimic the observational approach as closely as possible, we estimate it directly from the synthetic broadband luminosities. For each galaxy in the EAGLE-SKIRT catalogue, $L_{\text{TIR}}$ is calculated using the five-band recipe from \citet{2013MNRAS.431.1956G},
\begin{multline}
L_{\text{TIR}}
=
2.023\,L_{24} + 0.523\,L_{70} + 0.390\,L_{100} \\+ 0.577\,L_{160} + 0.721\,L_{250}
\label{LTIR}
\end{multline}
where all luminosities are expressed in $L_\odot$. The calibration coefficients in this formula were derived by fitting the total infrared luminosity obtained by integrating model SEDs between 3 and 1100~$\mu$m for a sample of nearby galaxies from the KINGFISH programme \citep{2011PASP..123.1347K}. We checked the accuracy of this recipe by comparing the result to the actual TIR luminosity obtained by integrating the SED between 3 and 1100~$\mu$m for all EAGLE galaxies from the Ref-100 and Recal-25 simulations, and found excellent agreement.

The resulting $z\leqslant0.1$ total infrared luminosity function is shown in the bottom left panel of Fig.~{\ref{EAGLE-LF-z00-01.fig}}. The data points correspond to the HerMES total infrared luminosity function as obtained by \citet{2016MNRAS.456.1999M}, where the total infrared was obtained by integrating the SED fits of the HerMES sources over the total infrared wavelength range.  The agreement between the EAGLE-SKIRT and HerMES luminosity functions is excellent over the entire luminosity range. 

\subsection{The dust mass function}
\label{DMF.sec}

The total dust mass is another fundamental characteristic for the interstellar medium in galaxies. It is a measure for the reservoir of metals that are locked up in grains, and combined with stellar masses and/or gas masses, it forms a powerful measure of the evolutionary stage of a galaxy \citep{2012A&A...540A..52C, 2017MNRAS.464.4680D, 2017MNRAS.471.1743D}. In the past few years, dust has also been used more frequently as a tracer for the total ISM mass budget \citep{2010A&A...518L..62E, 2012ApJ...761..168E, 2017ApJ...837..150S, 2017MNRAS.468L.103H}. As a result, many different studies have investigated the dependence of the total dust mass in galaxies as a function of galaxy type or environment \citep[e.g.,][]{2012A&A...540A..52C, 2016MNRAS.459.3574C, 2013MNRAS.428.1880A, 2014A&A...565A.128C, 2019A&A...626A..63D}, and some estimates for the dust mass function have been presented \citep{2000MNRAS.315..115D, 2011MNRAS.417.1510D, 2005MNRAS.364.1253V, 2015MNRAS.452..397C, 2018MNRAS.479.1077B}.

To determine an estimate of the dust mass, a simple modified blackbody model is fit to the luminosities in the PACS 160 $\mu$m and SPIRE 250, 350, and 500 $\mu$m bands. The parameters of this modified blackbody fitting are also available for all galaxies in the public EAGLE-SKIRT database \citep{2016A&C....15...72M, 2018ApJS..234...20C}. In the database, and in the post-processing itself, we have used the \citet{2004ApJS..152..211Z} BARE\_GR\_S dust model, characterised by a dust emissivity index $\beta=2$ and absorption coefficient $\kappa_{\text{abs}} = 0.057~{\text{m}}^2\,{\text{kg}}^{-1}$ at 850~$\mu$m. Most observational determinations of the dust mass function, however, are based on the MAGPHYS SED fitting code \citep{2008MNRAS.388.1595D}, which uses a dust model with the same emissivity index and $\kappa_{\text{abs}} = 0.077~{\text{m}}^2\,{\text{kg}}^{-1}$ at 850~$\mu$m \citep{2000MNRAS.315..115D, 2002MNRAS.335..753J}. In order to make our dust mass function directly comparable to observations, we have rescaled our EAGLE-SKIRT dust masses from the BARE\_GR\_S to the MAGPHYS scale by multiplying them with a factor $0.057/0.077= 0.74$.

The resulting EAGLE-SKIRT dust mass function for the redshift range $z\leqslant0.1$ is shown in the bottom right panel of Fig.~{\ref{EAGLE-LF-z00-01.fig}}. It is compared to the observational dust mass functions obtained by \citet{2011MNRAS.417.1510D} and \citet{2018MNRAS.479.1077B} for the same redshift range. Comparison to the latter work is an especially powerful probe, as it was based on a sample of more than 15,000 galaxies drawn from the combined GAMA \citep{2011MNRAS.413..971D} and H-ATLAS \citep{2010PASP..122..499E} surveys. The agreement between the EAGLE-SKIRT and GAMA/H-ATLAS dust mass function is nearly perfect for dust masses $M_{\text{d}}\lesssim2\times10^7~M_\odot$. Above this mass, the EAGLE-SKIRT dust mass function starts to underestimate the observed dust mass function, indicating a clear deficit of very dusty galaxies in the EAGLE simulation. This systematic underestimation is the direct translation of the underestimation of the luminosity functions in the SPIRE bands (Sect.~{\ref{LF-SPIRE.sec}}), on which the derived dust masses are based. 

\subsection{The infrared cosmic spectral energy distribution}
\label{CSED.sec}

The cosmic spectral energy distribution or CSED \citep{2008ApJ...678L.101D, 2012MNRAS.427.3244D, 2016MNRAS.455.3911D} represents the total electromagnetic power generated within a cosmological unit volume as a function of wavelength. Being a complex function of both the volume density of different galaxy populations and the different processes that shape the SED of a single galaxy, it is a fundamental observational characteristic of the Universe. In \citet{2019MNRAS.484.4069B} we used the EAGLE-SKIRT database to generate the UV--submm CSED of the EAGLE simulation, and compared it to the observed GAMA CSED \citep{2017MNRAS.470.1342A}. Except in the UV, where the EAGLE-SKIRT CSED overestimated the observed values by up to an order of magnitude, we found an excellent agreement between the observed and simulated CSED in the Local Universe. At infrared wavelengths, the agreement was particularly satisfactory \citep[see][Fig.~1]{2019MNRAS.484.4069B}, especially when the EAGLE-SKIRT CSED was compared to the HerMES CSED presented by \citet{2016MNRAS.456.1999M}. Still, the EAGLE-SKIRT CSED systematically underestimates the observed HerMES CSED. The difference is small, typically below 0.1 dex, but systematic.

Different reasons were put forward to explain this minor discrepancy, namely the insensitivity to luminous sources because of the small volume probed by the Recal-25 simulation, and the underestimation of the UV attenuation in the star-forming regions. One additional aspect that was not considered in detail is the difference in methodology to compute the CSED. We calculated the EAGLE-SKIRT CSED in a very simple, straightforward way: at every wavelength, we simply summed the observed luminosities of every single galaxy in the EAGLE-SKIRT database, and subsequently normalised the sum based on the snapshot co-moving volume. Observed CSEDs are usually calculated in a more complex two-step way \citep[e.g.,][]{2012MNRAS.427.3244D, 2016MNRAS.455.3911D, 2017MNRAS.470.1342A}. First the luminosity function is calculated at every wavelength, and subsequently, this luminosity function, or rather the parameterised fit to it, is integrated over the entire luminosity range. This approach has the advantage that it can take into account the contribution of the low- and high-luminosity tails of the distribution. On the other hand, these contributions can be uncertain, as they are based on extrapolations of analytical fits to a limited number of observed data points.

With the EAGLE-SKIRT infrared luminosity functions we have derived, we can now mimic more closely the observational methodology to calculate the EAGLE-SKIRT infrared CSED. Fig.~{\ref{CSED-z00-01.fig}} shows the comparison of the $z\leqslant0.1$ CSED obtained in this way with the CSED presented by \citet{2019MNRAS.484.4069B}. Also shown on this plot is the observed HerMES CSED, as presented by \citet{2016MNRAS.456.1999M}. It is clear that the results of the two methods are completely compatible, which implies that the conclusions drawn by \citet{2019MNRAS.484.4069B} are still fully valid.

\section{Cosmic evolution}
\label{LFevo.sec}

\begin{figure*}
\includegraphics[width=0.49\textwidth]{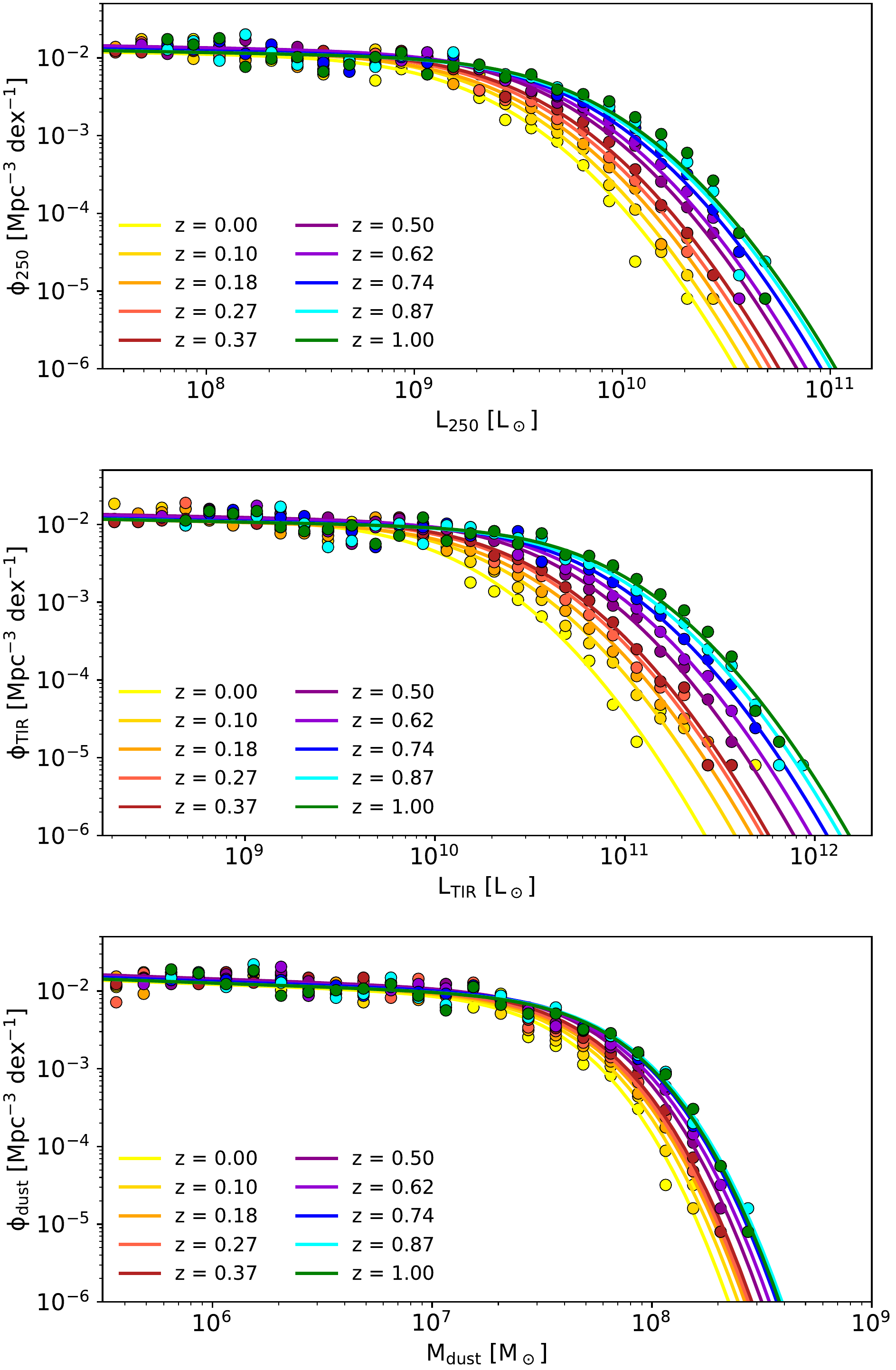}
\includegraphics[width=0.49\textwidth]{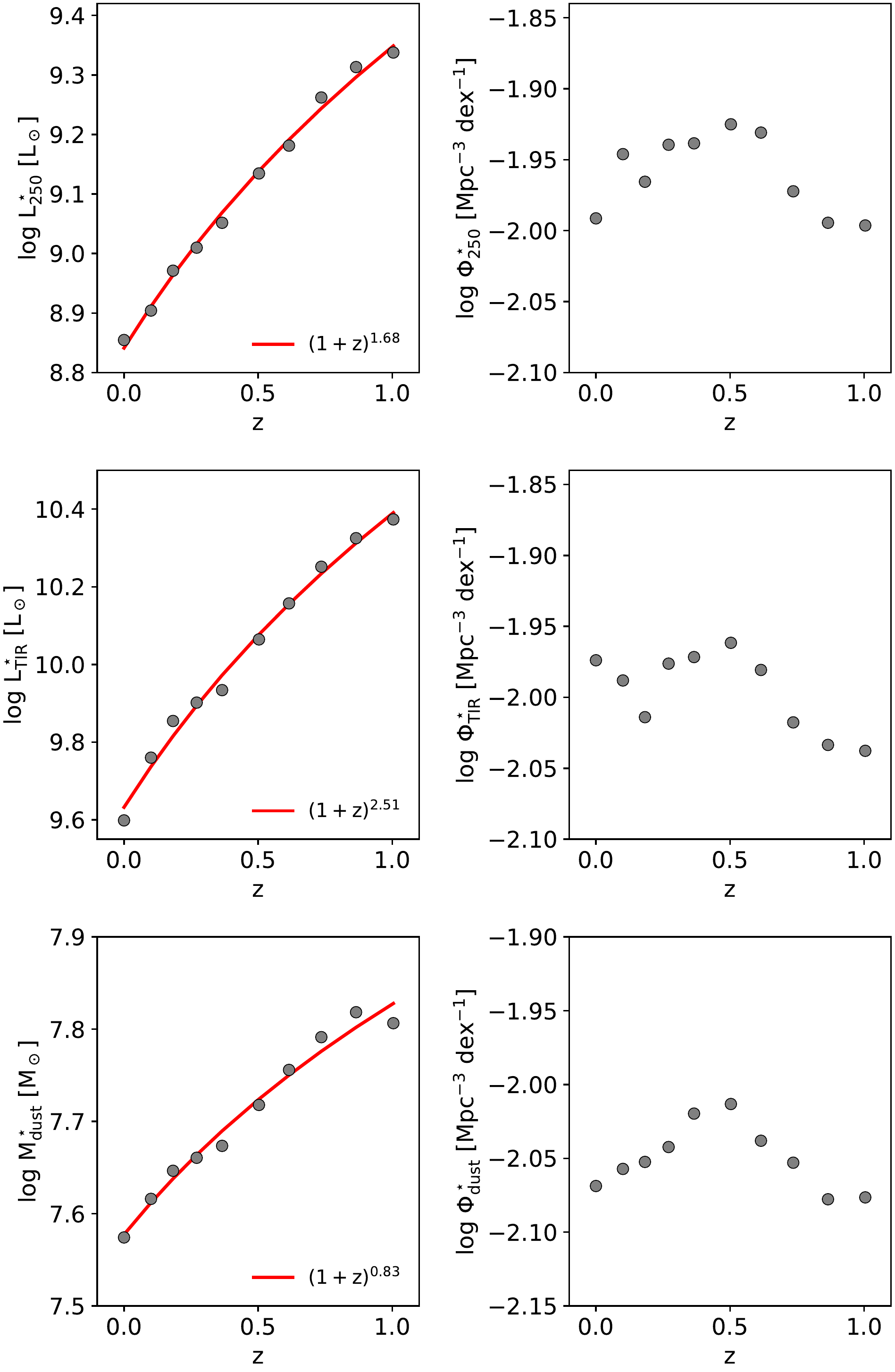}
\caption{Evolution of the EAGLE-SKIRT 250 $\mu$m luminosity function (top row), total infrared luminosity function (middle row) and dust mass function (bottom row) from $z=0$ to $z=1$. The left panel on each row shows the actual luminosity/mass functions for the different EAGLE snapshots, with each colour corresponding to a different redshift. The dots represent the luminosity function as obtained from the EAGLE-SKIRT database, and the solid line is the best modified Schechter fit to these data points. The middle and right panels on each row show the explicit redshift evolution of the characteristic luminosity and density, as derived from these modified Schechter fits.} 
\label{Evolution-LFs.fig}
\end{figure*}

\subsection{Luminosity functions}
\label{LFevoevo.sec}

Observations have indicated clear evidence for strong evolution of the infrared luminosity functions over the past few Gyr. At infrared wavelengths, IRAS, ISO and Spitzer surveys revealed indications of rapid evolution \citep[e.g.,][]{1990MNRAS.242..318S, 2001MNRAS.325..665C, 2004MNRAS.355..813S, 2004ApJ...609..122P, 2005ApJ...632..169L, 2006MNRAS.370.1159B, 2007ApJ...660...97C, 2010A&A...515A...8R}. All of these studies seem to exclude scenarios that favour a larger evolution in density than in luminosity. In other words, these studies point to a rapid evolution of the characteristic luminosity $L^\star$ of the best-fit modified Schechter function, rather than a strong evolution of the normalisation factor $\Phi^\star$.

At submm wavelengths, tentative evidence for similar evolution was provided by \citet{2009ApJ...707.1779E}, based on BLAST balloon observations of the GOODS-South field: the data suggest strong evolution out to $z = 1$ in both monochromatic luminosity function and dust mass function, particularly among the higher luminosity/mass systems. One of the first results of the Herschel mission was a confirmation of this tentative evidence. Based on the first 14 deg$^2$ of extragalactic sky observed in the frame of the H-ATLAS survey, \citet{2010A&A...518L..10D} determined the luminosity function of 250 $\mu$m-selected galaxies out to $z=0.5$, and clearly demonstrated steady evolution out to this redshift. Based on the HerMES data, \citet{2016MNRAS.456.1999M} reinforces this evidence: their luminosity functions show significant and rapid luminosity evolution already at redshifts at low as $z\leqslant0.2$. \citet{2011MNRAS.417.1510D} presented evidence for a strong evolution of the dust mass function out to $z=0.5$. 

At higher redshifts, \citet{2010A&A...518L..23E} used the first HerMES data to investigate the evolution of the SPIRE 250 $\mu$m luminosity function out to $z=2$. They found strong evolution out to $z\sim1$, but no or at most weak evolution between $z\sim1$ and $z\sim2$. Based on PACS and SPIRE data from the PACS Evolutionary Probe \citep[PEP:][]{2011A&A...532A..90L} and HerMES surveys, respectively, \citet{2013MNRAS.432...23G} reported very strong luminosity evolution to $z\sim2$, and milder evolution to larger redshifts.

The left panels of Fig.~{\ref{Evolution-LFs.fig}} show the 250 $\mu$m and total infrared luminosity functions\footnote{To calculate the total infrared luminosities of the EAGLE sources, we apply the five-band recipe from \citet{2013MNRAS.431.1956G} on the EAGLE-SKIRT rest-frame luminosities. While this formula was calibrated on local sources, it gives a reliable measure of $L_{\text{TIR}}$ for higher-redshift galaxies as well. We have checked this by comparing the recipe to the actual TIR luminosities for EAGLE galaxies at $z=0.5$ and $z=1$.}, as well as the dust mass function, for 10 different redshifts between $z=0$ and $z=1$. These 10 redshifts correspond to the last 10 snapshots of the EAGLE simulations. For each luminosity/mass function, we have fixed the value of $\alpha$ and $\sigma$ to the mean value of all redshift slices ($\alpha_{250} = 1.05$, $\sigma_{250} = 0.40$, $\alpha_{\text{TIR}} = 1.05$, $\sigma_{\text{TIR}} = 0.43$, $\alpha_{\text{dust}} = 1.10$, $\sigma_{\text{dust}} = 0.20$), and hence just used the characteristic luminosity $L^\star$ and density $\Phi^\star$ as free parameters. In other words, we allow for both luminosity/mass and density evolution of the luminosity/mass functions. Note that we fitted the data points at each redshift independently, and we did not build in a parameterised redshift evolution, as is sometimes done \citep[e.g.,][]{2005ApJ...632..169L, 2013MNRAS.428..291P}. These panels immediately show significant evolution for the infrared luminosity functions, and a more moderate evolution in the dust mass function.

To quantify this evolution, we show on the middle and right panels of Fig.~{\ref{Evolution-LFs.fig}} the variation of the characteristic luminosity/mass and density explicitly as a function of redshift, up to $z=1$. The grey dots represent the fitted parameter values for each individual luminosity/mass function, whereas the red lines show fits to these data points of the standard form,
\begin{gather}
L^\star(z) \propto (1+z)^{\alpha_{\text{L}}}, \label{Lstar} \\ 
M^\star(z) \propto (1+z)^{\alpha_{\text{M}}}, \label{Mstar}
\end{gather}
Interestingly, we find a combination of relatively mild luminosity evolution for the 250 $\mu$m and TIR luminosity functions ($\alpha_{\text{L,250}}\sim1.68$ and $\alpha_{\text{L,TIR}}\sim2.51$) and very limited density evolution. Actually, the 250~$\mu$m and total infrared luminosity function are compatible with zero density evolution and hence pure luminosity evolution. For the dust mass function, there is a mild mass evolution ($\alpha_{\text{M}}\sim0.83$) and again, no strong evidence for density evolution.

It is interesting to compare these results to those obtained observationally. For the 250~$\mu$m band, \citet{2016MNRAS.456.1999M}, based on a very limited redshift range out to $z=0.15$, report a very strong luminosity evolution ($\alpha_{\text{L},250} = 5.3\pm0.2$) combined with a mild negative density evolution ($\alpha_{\text{D},250} = -0.6\pm0.4$). \citet{2016A&A...592L...5W} studied the evolution of the 250 $\mu$m luminosity function down to much fainter luminosities using a modified stacking method. They also find a combination of strong positive luminosity evolution ($\alpha_{\text{L},250} = 4.89\pm1.07$) and moderate negative density evolution ($\alpha_{\text{D},250} = -1.02\pm0.54$) over the redshift range $0.02\leqslant z\leqslant0.5$.

For the total infrared luminosity function, \citet{2016MNRAS.456.1999M} report a very rapid luminosity evolution, combined with a significant negative density evolution ($\alpha_{\text{L,TIR}} = 6.0\pm0.4$ and $\alpha_{\text{D,TIR}} = -2.1\pm0.4$). This results are at odds with those obtained by other teams. Based on MIR data out to 24~$\mu$m and the assumption of a quasi-linearity between the monochromatic mid-infrared luminosity at 15 $\mu$m and the total infrared luminosity, \citet{2005ApJ...632..169L} calculate the evolution of the TIR galaxy luminosity function out to $z\sim1$. Their best fit corresponds to a combination of positive evolution in both density and luminosity ($\alpha_{\text{L,TIR}} = 3.2^{+0.7}_{-0.2}$ and $\alpha_{\text{D},250} = 0.7^{+0.2}_{-0.6}$), although their data is also compatible with strict evolution in luminosity and no density evolution. Also \citet{2010A&A...515A...8R} report evidence for a relatively mild luminosity evolution combined with a positive density evolution ($\alpha_{\text{L,TIR}} \sim 2.7$ and $\alpha_{\text{D,TIR}} \sim 1.1$) over the redshift range $z\leqslant1$.

Given this spread in literature results, our pure luminosity evolution seems to be a fairly reasonable middle ground. Concerning the rate of the luminosity evolution, however, it seems that the EAGLE-SKIRT values that our calculations reveal ($\alpha_{\text{L}}\sim2$) are on the low side, in agreement with the CSED results obtained by \citet{2019MNRAS.484.4069B}. 

\subsection{Infrared luminosity density and dust mass density}

\begin{figure}
\includegraphics[width=0.95\columnwidth]{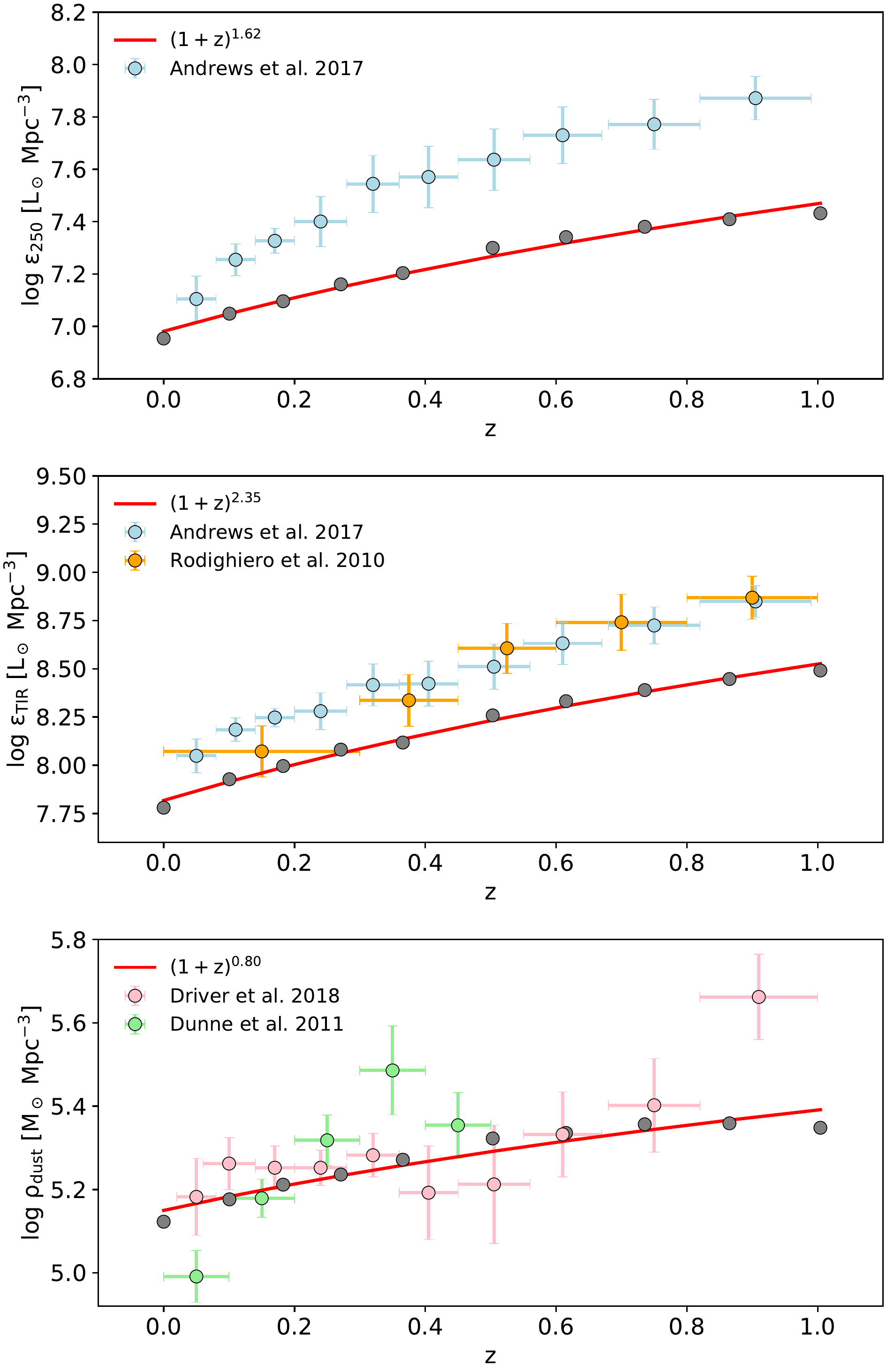}
\caption{Redshift evolution of the 250~$\mu$m luminosity density, the TIR luminosity density, and the dust mass density, as obtained from the EAGLE-SKIRT luminosity functions and dust mass functions shown in Fig.~{\ref{Evolution-LFs.fig}}. The grey dots are the parameter values as obtained from the modified Schechter fits, the red lines are fits to these data points. The data points with error bars represent observational values from different sources, as indicated in the top left corner of each panel.} 
\label{Evolution-rho.fig}
\end{figure}

In Fig.~{\ref{Evolution-rho.fig}} we plot the evolution of the 250~$\mu$m luminosity density $\varepsilon_{250}$, the total infrared luminosity density $\varepsilon_{\text{TIR}}$, and the dust mass density $\rho_{\text{dust}}$, explicitly as a function of redshift. As a result of the mild luminosity/mass evolution and the absence of significant density evolution found for the luminosity/mass functions, we also obtain a relatively mild evolution of these quantities. 

In the top panel, we also show observational data points corresponding to the GAMA CSED evolution study by \citet{2017MNRAS.470.1342A}. Note that these data do not correspond to the data in Tables~1 and 2 of their paper, however, as these correspond to the observed reference frame. Instead, we used the MAGPHYS fits to the CSED at different redshifts as provided online by \citet{2017MNRAS.470.1342A}, and convolved these fits with the SPIRE 250~$\mu$m transmission curves. The EAGLE-SKIRT results do not reproduce the evolution seen in the GAMA data: the evolution in EAGLE-SKIRT is far too modest, which leads to a rapid build-up of a systematic underestimation of $\sim0.4$~dex. This finding is in agreement with our previous results on the CSED \citep{2019MNRAS.484.4069B}. Indeed, we found that EAGLE-SKIRT increasingly underestimates the observed GAMA CSED, with the largest discrepancies found at far-infrared and submm wavelengths. 

The middle panel of Fig.~{\ref{Evolution-rho.fig}} shows the evolution of the total infrared luminosity density. We compare our EAGLE-SKIRT results to the observational values by \citet{2010A&A...515A...8R} and \citet{2017MNRAS.470.1342A}. For the latter study, we have determined the total infrared luminosity density by formula (\ref{LTIR}) on the monochromatic luminosity densities obtained from the MAGPHYS fits, as described above. The two observational data sets agree fairly well, especially at $z\gtrsim0.5$. At the lowest redshifts, we believe that the \citet{2017MNRAS.470.1342A} might be slightly overestimated, due to the peculiar shape of the MAGPHYS fits between 24 and 100~$\mu$m \citep[see discussion in Sect.~3.5 of][]{2019MNRAS.484.4069B}. The EAGLE-SKIRT data clearly underestimate the rapid evolution in the total infrared luminosity density: we obtain an evolution as $(1+z)^{2.3}$, where \citet{2005ApJ...632..169L} and \citet{2010A&A...515A...8R} report a much stronger evolution as $(1+z)^{3.9\pm0.4}$ and $(1+z)^{3.8\pm0.4}$, respectively. At $z\sim1$, our EAGLE-SKIRT postprocessing recipe underestimates the observed total infrared luminosity density by $\sim0.5$~dex.

Finally, in the bottom panel of Fig.~{\ref{Evolution-rho.fig}}, we compare the evolution of the EAGLE-SKIRT dust mass density with observational estimates by \citet{2011MNRAS.417.1510D} and \citet{2018MNRAS.475.2891D}. Interestingly, the data sets, both based on MAGPHYS modelling, seem to be incompatible. The measurements by \citet{2011MNRAS.417.1510D} are based on nearly 2000 SPIRE-selected sources from the H-ATLAS SDP data \citep{2011MNRAS.415.2336R}. Ignoring their last data point, which they argue is prone to incompleteness and/or photo-$z$ bias, \citet{2011MNRAS.417.1510D} find a very strong evolution in the dust mass density out to $z\sim0.4$, which can be well described by $\rho_{\text{dust}}\propto(1+z)^{4.5}$. The dust mass density evolution shown by \citet{2018MNRAS.475.2891D} is based on roughly half a million sources from the GAMA, G10-COSMOS \citep{2015MNRAS.447.1014D, 2017MNRAS.464.1569A} and 3D-HST \citep{2012ApJS..200...13B} surveys.  They do not recover the steep evolution at low redshifts found by \citet{2011MNRAS.417.1510D}: instead, they report a relatively flat dust mass density function. Except for their data point at $z\sim0.9$, our EAGLE-SKIRT results reproduce the \citet{2018MNRAS.475.2891D} data fairly well.

\section{Discussion}
\label{Discussion.sec}

The main conclusion from this paper is twofold. On the one hand, we can conclude that the EAGLE simulation, or more precisely the EAGLE-SKIRT post-processing recipe to generate synthetic multi-wavelength observations for the EAGLE galaxies, reproduces the infrared luminosity and dust mass functions in the local Universe ($z\leqslant0.2$) very well. Both the shape and the normalisation of the luminosity function are recovered well in nearly all infrared bands considered. Some minor discrepancies are encountered, mainly in the high luminosity regime, where the EAGLE-SKIRT luminosity functions mildly but systematically underestimate the observed ones. A very important result is the excellent agreement between EAGLE-SKIRT and observations for the total infrared luminosity function. While some discrepancies in dust-related scaling relations \citep{2017MNRAS.470..771T, Trcka2019} point to imperfections in the details of the dust absorption and re-emission of our EAGLE-SKIRT framework, this agreement shows that the global energy budget for attenuation in this framework is appropriate.

\begin{figure}
\includegraphics[width=0.93\columnwidth]{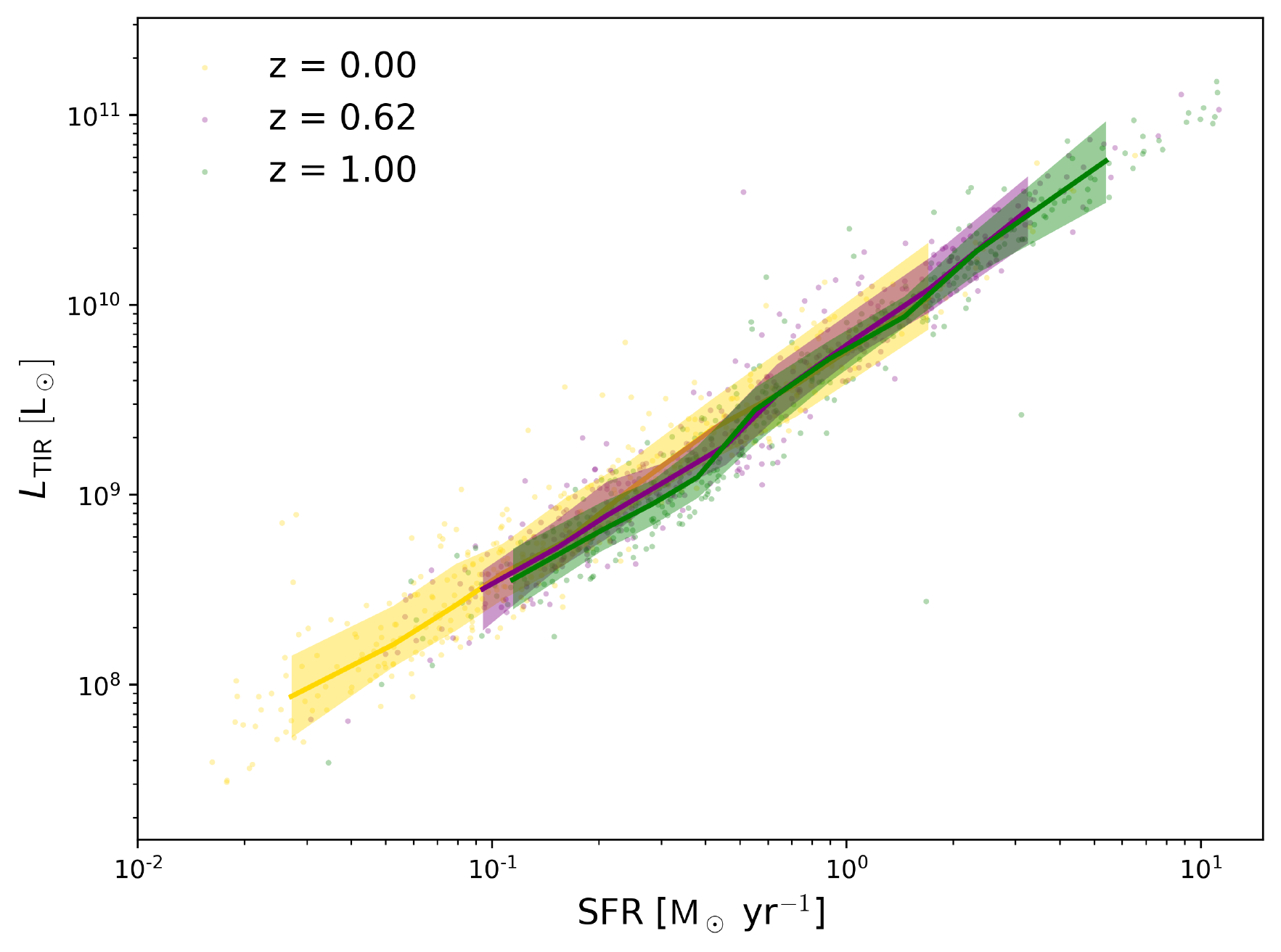}
\caption{Correlation between SFR and the total infrared luminosity for galaxies from the EAGLE Recal-25 simulation. The different colours correspond to galaxies at three different redshifts between 0 and 1. The solid lines indicate the running median, and the shaded regions indicate the 20-80\% percentile zones.}
\label{SFRvsL.fig}
\end{figure}

On the other hand, the agreement between the EAGLE-SKIRT infrared luminosity functions and the observed ones gradually worsens with increasing redshifts. Fitting modified Schechter functions to the EAGLE-SKIRT luminosity and dust mass functions at different redshifts, we find a combination of relatively mild luminosity evolution and very limited density evolution for the 250 $\mu$m and TIR luminosity functions out to $z=1$. The 250 $\mu$m and total infrared luminosity functions are compatible with zero density evolution and hence pure luminosity evolution. For the dust mass function, we find a mild mass evolution and again, no strong evidence for density evolution. The results in the literature concerning the nature of the evolution of the luminosity function are diverse, with some teams advocating very strong luminosity evolution combined with negative density evolution \citep{2016MNRAS.456.1999M}, and other teams finding a milder luminosity evolution and a mild positive density evolution \citep{2005ApJ...632..169L, 2010A&A...515A...8R}. In any case, concerning the rate of luminosity evolution, our EAGLE-SKIRT results are on the low side, and our predicted evolution of the infrared luminosity density is significantly weaker than the observed trends. 

Compared to the luminosity density, our estimate for the evolution of the cosmic dust mass density is in fairly good agreement with observational data, especially with the recent evolutionary trends derived from GAMA, G10-COSMOS and 3D-HST observations \citep{2018MNRAS.475.2891D}. This agreement is interesting, because there could be various reasons why one could expect an increasing disagreement with increasing redshift \citep[see also the discussion in][]{2019MNRAS.484.4069B}. Both the subgrid physics in the EAGLE simulations \citep{2015MNRAS.450.1937C} and the EAGLE-SKIRT post-processing radiative transfer procedure \citep{2016MNRAS.462.1057C, 2018ApJS..234...20C} are calibrated against observed relations in the local Universe, and hence are not necessarily optimised for the higher redshift Universe. In particular, a single set of dust optical properties and a single dust-to-metal ratio were adopted at all redshifts, while it is optimistic to assume that this simple recipe is realistic. Given these uncertainties, the correspondence in the bottom panel of Fig.~{\ref{Evolution-rho.fig}} is very encouraging. 

It might seem odd at first that we recover the evolution of the dust mass density relatively well, whereas we significantly and systematically underestimate the evolution of the infrared luminosity density. This suggests that the EAGLE-SKIRT procedure allocates roughly the correct amount of dust in each galaxy, but that this dust underperforms more and more in absorbing and re-emitting starlight with increasing redshift. Part of this discrepancy is probably due to the inherent resolution limitations of the radiative transfer process. A poor spatial resolution is a significant threat for reliable results from radiative transfer simulations, and it generally leads to an underestimation of the absorption and re-emission efficiency \citep[e.g.,][]{2015A&A...576A..31S, 2018A&A...616A.120M}. As the simulated EAGLE galaxies at increasing redshifts have smaller masses and hence fewer particles, it can be expected that these resolution effects increase with increasing look-back time. Higher resolution cosmological simulations would be welcome to test the importance of this effect.

Secondly, we believe that limitations in the EAGLE simulations themselves could also contribute to this discrepancy. The TIR luminosity is a well-known proxy for the SFR in galaxies \citep{2009ApJ...692..556R, 2012ARA&A..50..531K}. In Fig.~{\ref{SFRvsL.fig}} we show the total infrared luminosity as a function of SFR for all EAGLE Recal-25 galaxies at three different redshifts between 0 and 1. Interestingly, the strong correlation between $L_{\text{TIR}}$ and SFR appears to be essentially independent of redshift, at least up to $z=1$. This suggests that the evolution of the infrared luminosity functions is strongly connected to the evolution of the SFR function. \citet{2017MNRAS.472..919K} presented the evolution of the SFR function and the SFR density for the EAGLE simulations. They find a general underestimation of the SFR density with respect to observational estimates, and this underestimation increases with increasing redshift (see their Fig.~3a). \citet{2018MNRAS.473.3507E} also noted that the EAGLE SFR function seems to show a slower evolution than the observed one, and the same effect may be reflected in the integrated and resolved star forming main sequences \citep{2015MNRAS.450.4486F, 2019MNRAS.485.5715T}. This too mild evolution in SFR density will also contribute to the too mild evolution in infrared luminosity density, even though the cosmic dust density is reproduced fairly well. 

Finally, we reiterate the caveat that our EAGLE-SKIRT catalogue misses some infrared-bright sources, which do contribute to the observed infrared luminosity functions and infrared luminosity density. In particular, the infrared emission by AGNs is not included in our radiative transfer post-processing recipes. With the AGN density a strong function of redshift \citep{2005A&A...441..417H, 2009MNRAS.399.1755C, 2011ApJ...728...56A}, this might also contribute substantially to the growing disagreement between our EAGLE results and observations with increasing look-back time.

\section{Summary}
\label{Summary.sec}

We have presented infrared luminosity functions and dust mass functions for the EAGLE cosmological simulation. These luminosity functions and dust mass functions are based on synthetic infrared luminosities, generated by means of the EAGLE-SKIRT post-processing recipe presented by \citet{2016MNRAS.462.1057C, 2018ApJS..234...20C}. The goal of this paper was to compare these EAGLE-SKIRT luminosity and dust mass functions to observational ones, as a test of the EAGLE simulations and the EAGLE-SKIRT recipe. The main results of this paper are the following:
\begin{itemize}
\item
In the local Universe ($z\leqslant0.2$), we reproduce the observed infrared luminosity functions very well: both the shape and the normalisation are recovered well in nearly all infrared bands considered. Minor deviations are found, primarily at the high-luminosity tail of the luminosity functions. We argue that hese differences are due to a combination of factors, including imperfections in the EAGLE-SKIRT calibration procedure, the lack of AGN and lensing effects in our analysis, and the onset of cosmic evolution.
\item 
We reproduce the shape and normalisation of the total infrared luminosity and dust mass function in the local Universe. This shows that the global energy budget for attenuation in the EAGLE-SKIRT framework is appropriate, even though some discrepancies in dust-related scaling relations point to imperfections in the details of the dust absorption and re-emission \citep{2017MNRAS.470..771T, Trcka2019}.
\item
We use modified Schechter fits to the luminosity functions to calculate the infrared CSED in a way that mimics the observational methodology. The resulting values are in good agreement with those obtained by \citet{2019MNRAS.484.4069B} in a simpler way.
\item 
We study the evolution of the EAGLE-SKIRT infrared luminosity functions and dust mass functions out to $z=1$. We quantify the evolution by fitting modified Schechter functions to the luminosity/mass functions at different redshifts, and by subsequently investigating the evolution of the best-fit parameters. These fits yield a relatively mild luminosity evolution, combined with no or very limited density evolution for the infrared luminosity and dust mass functions. Concretely, we find an evolution of $L_{\star,250}\propto(1+z)^{1.68}$, $L_{\star,\text{TIR}}\propto(1+z)^{2.51}$ and $M_{\star,\text{dust}}\propto(1+z)^{0.83}$.
\item
We find a dust mass density evolution of $\rho_{\text{dust}} \propto (1 + z)^{0.80}$ out to $z=1$. This evolution is in reasonable agreement with the one derived from GAMA, G10-COSMOS and 3D-HST observations \citep{2018MNRAS.475.2891D}.  On the contrary, the evolution of the EAGLE-SKIRT infrared luminosity densities underestimates the observed evolution significantly and systematically. For the 250 $\mu$m and total infrared luminosity density we find modest evolutions of $\varepsilon_{250} \propto (1 + z)^{1.62}$ and $\varepsilon_{\text{TIR}} \propto (1 + z)^{2.35}$, weaker than observational estimates. These differences can be due to a combination of different factors: the radiative transfer calculations might become increasingly inaccurate because of the increasingly poor resolution, the evolution of the SFR density in the EAGLE simulation seems to be slower than the observed one, and we miss the contribution of infrared emission by AGNs in our EAGLE-SKIRT post-processing recipe. 
\end{itemize}

\section*{Acknowledgements}
MB, AT, PC and BV gratefully acknowledge financial support from the Fund for Scientific Research Flanders (FWO-Vlaanderen, project G.0392.16N), and from the Belgian Science Policy Office (BELSPO, PRODEX Experiment Arrangement project C4000128500). The EAGLE-SKIRT database on which this work was based used the DiRAC Data Centric system at Durham University, operated by the Institute for Computational Cosmology on behalf of the Science and Technology Facilities Council (STFC) DiRAC High Performance Computing (HPC) Facility ({\href{http://www.dirac.ac.uk}{http://www.dirac.ac.uk}}). This equipment was funded by BIS National E-infrastructure capital grant ST/K00042X/1, STFC capital grants ST/H008519/1 and ST/K00087X/1, STFC DiRAC Operations grant ST/K003267/ 1, and Durham University. DiRAC is part of the National E- Infrastructure.

\bibliographystyle{mnras}
\bibliography{EAGLE}

\label{lastpage}
\end{document}